\documentclass[a4paper,11pt]{article}
\usepackage{geometry}
\usepackage[svgnames]{xcolor}
\usepackage[colorlinks=true,urlcolor =black,linkcolor=blue,citecolor=blue,bookmarks=false]{hyperref}
\usepackage{amsfonts,amsmath,amssymb}
\usepackage{graphicx}
\usepackage{dsfont}
\usepackage{graphicx}
\usepackage{bm}
\usepackage[font=small,labelfont=bf]{caption}

\usepackage{enumitem}
\usepackage{boldline}
\usepackage{changepage}
\usepackage{cite}
\usepackage{textcomp}
\usepackage[title]{appendix}
\usepackage{url}
\usepackage[OT1]{fontenc}
\usepackage{authblk}
\usepackage{bbold}

\usepackage{etoolbox}
\makeatletter
\patchcmd{\l@section}
  {\hfil}
  {\leaders\hbox{\normalfont$\m@th\mkern \@dotsep mu\hbox{.}\mkern \@dotsep mu$}\hfill}
\makeatother

\title{\bf Transitions between quasi-stationary states in traffic systems: Cologne orbital motorways as an example }

\author{Shanshan Wang \thanks{shanshan.wang@uni-due.de}, Michael Schreckenberg and Thomas Guhr}
\affil{\textit{Faculty of Physics, University of Duisburg--Essen, Duisburg, Germany}}

\date{\today}
 
\begin{document}
\maketitle

\noindent {\bf Abstract.}
Traffic systems can operate in different modes. In a previous work, we identified these modes as different quasi-stationary states in the correlation structure. Here, we analyze the transitions between such quasi-stationary states, i.e., how the system changes its operational mode. In the longer run this might be helpful to forecast the time evolution of correlation patterns in traffic. We take Cologne orbital motorways as an example, we construct a state transition network for each quarter of 2015 and find a seasonal dependence for those quasi-stationary states in the traffic system. Using the PageRank algorithm, we identify and explore the dominant states which occur frequently within a moving time window of 60 days in 2015. To the best of our knowledge, this is the first study of this type for traffic systems. 

\vspace{0.5cm}
\noindent{\bf Keywords\/}: quasi-stationary state, traffic system, transition probability matrix, PageRank, reduced-rank correlation matrix
\vspace{1cm}

\noindent\rule{\textwidth}{1pt}
\vspace*{-1cm}
{\setlength{\parskip}{0pt plus 1pt} \tableofcontents}
\noindent\rule{\textwidth}{1pt}

\section{Introduction}
\label{sec1}
Time series of flows and velocities in traffic systems exhibit fluctuations that indicate and reflect non-stationarity. Nevertheless, traffic systems organize themselves in different operational modes that exist for certain time spans. Put differently, there are quasi-stationary states. Recently, we identified them in the correlation structures~\cite{Wang2020}, transferring approaches from the study of financial markets~\cite{Munnix2012,Chetalova2015a,Chetalova2015b,Guhr2015,Rinn2015,Stepanov2015,Heckens2020,Pharasi2020a,Pharasi2020b,Pharasi2021}. To understand the system dynamics, transitions between these states are of particular interest. In finance this issue has been studied in references~\cite{Munnix2012,Chetalova2015a,Chetalova2015b,Guhr2015,Rinn2015,Stepanov2015,Heckens2020,Pharasi2020a,Pharasi2020b,Pharasi2021}. Non- and quasi-stationarity has also been studied with different observables, e.g.,  in climate~\cite{Mann2004,Cheng2014}, power systems~\cite{Messina2009}, speech recognition~\cite{Cohen2001,Rangachari2006} and in other systems. 

Most studies on traffic systems focus on modeling and simulation either from macroscopic or from microscopic perspectives~\cite{Nagel1992,Schadschneider1993,Lovaas1994,Schreckenberg1995,Hoogendoorn2001,Wong2002,Fellendorf2010,Treiber2013}. Empirical studies~\cite{Kerner2002,Bertini2005,Schonhof2007,Kerner2012} are less frequent due to the limited amount of available traffic data. In a previous study~\cite{Wang2020}, we classified the traffic system of Cologne orbital motorways in Germany into six quasi-stationary states in the temporal correlation structure of the traffic flow. How the six states evolve with time and which state dominates in a traffic system remains to be addressed. Progress will be helpful for understanding the evolution of traffic behavior and forecasting purposes. We aim at studying the transition between quasi-stationary states, taking Cologne orbital motorways as an example. We work out the transition probabilities between states within time windows. Using the transition probability matrices together with the PageRank algorithm~\cite{Brin1998,Page1999,Langville2011}, we identify the dominant states in the traffic system and analyze the influence of random events on the dominant states. We also explore the number of transition steps that the system typically takes to return to a state.   

The PageRank algorithm was first developed for the Google search engine to determine which webpages are important based on their link structures~\cite{Brin1998,Page1999,Langville2011}. The link structure of webpages can be formulated mathematically in a transition matrix, and the webpages are modeled as the nodes of a graph constructed from the transition matrix. To ensure that the resulting importance scores, referred to as the PageRank score, and uniquely exists, an external influence on the importance of each node is introduced~\cite{Boldi2009}. Due to its simplicity, generality, guaranteed existence, uniqueness and computational fastness~\cite{Gleich2015}, PageRank has been extended to many fields, including biology~\cite{Morrison2005}, chemistry~\cite{Mooney2012}, neuroscience~\cite{Crofts2011}, mathematics~\cite{Frahm2012}, economics~\cite{Ermann2015, Coquide2019, Coquide2020}, etc. In particular, based on road and urban space networks, PageRank is also used to predict traffic flow of individual roads~\cite{Jiang2008} and to study Markov chain models for road planning and optimal routing~\cite{Schlote2012}. In our study, we utilize the PageRank algorithm to estimate the importance of each quasi-stationary state to identify dominant states in the studied traffic system. Moreover, we also employ the PageRank algorithm to explore the influence of random events on the dominant states.

The paper is organized as follows. In Sec.~\ref{sec2}, we identify the quasi-stationary states by performing $k$-means clustering for reduce-rank correlation matrices. In Sec.~\ref{sec3}, we describe the PageRank algorithm, visualize the state transition networks for four quarters of 2015 based on transition probability matrices, identify the dominant states with the PageRank algorithm, explore the influence of random events on the dominant states, and analyze the recurrence of quasi-stationary states. We finally discuss our results in Sec.~\ref{sec4}. 

\section{Identifying quasi-stationary states}
\label{sec2}
We developed the technique of analyzing correlation structures relative to the dominating collective motion and applied it in various studies~\cite{Heckens2020,Wang2020,Wang2022,Heckens2022a,Heckens2022b}. Related applications have been put forward in Refs.~\cite{Munnix2012,Chetalova2015a,Chetalova2015b,Guhr2015,Rinn2015,Stepanov2015,Pharasi2020a,Pharasi2020b,Pharasi2021,Wang2021}. However, in the cited works, the correlations of time series were the object of interest or, equivalently the correlation matrices of the positions, when the time series were measured. Here, we address the correlations of the position series, i.e., the corresponding correlations matrices are those of the times. For the sake of clarity and for the convenience of the reader, we thus briefly sketch the construction for reduced-rank correlation matrices in Sec.~\ref{sec21}, of course in the proper form for the correlations of times. We identify quasi-stationary states in the traffic system of Cologne orbital motorways by performing $k$-means clustering for reduced-rank correlation matrices in Sec.~\ref{sec22}.

\subsection{Reduced-rank correlation matrices}
\label{sec21}

The traffic data used in this study are from inductive loop detectors of $K=35$ available sections on the Cologne orbital motorways in Germany, consisting of part of the motorways A1, A3 and A4. The data include the traffic flows and velocities for each lane and each motorway section. We aggregate traffic flows across multiple lanes of a motorway section during a time period of 15 minutes. For 35 sections, we have 35 time series $F_k(t)$, $k=1,\cdots,K$, of aggregated traffic flow with the length of $T=96$ data points each day. All $F_k(t)$ enter into a $K\times T$ data matrix $F$. Importantly, we do not focus on the spatial correlations which are encoded in the time series $F_k(t),~t=1,\cdots,T$. Rather, we are interested in the mutual temporal dependencies that can be analyzed by correlating the position series $F_k(t),~k=1,\cdots,K$. Each column of $F$ is a position series that describes the traffic flows at different positions $k=1,\cdots,K$, but at the same time $t$. We normalize each position series to zero mean by 
\begin{equation}
A_{k}(t)=F_{k}(t)-\langle F_{k}(t) \rangle_K \ ,
\label{eq2.1.1}
\end{equation}
to obtain a new $K\times T$ data matrix $A$, where $\langle \cdots\rangle_K$ indicates the average over all positions $k$. The normalized data matrix $A$ yields a $T\times T$ covariance matrix of the position series by 
\begin{equation}
\Sigma=\frac{1}{K}A^{\dag}A \ .
\label{eq2.1.2}
\end{equation} 
The subscript $\dag$ represents the transpose of a matrix.  We carry out a spectral decomposition of the covariance matrix
\begin{equation}
\Sigma=\sum_{t=1}^{T}\Theta_tV_tV^{\dag}_t \ ,
\label{eq2.1.3}
\end{equation}
where $\Theta_t$ is the $t$-th eigenvalue of $\Sigma$, and $V_t$ is the corresponding eigenvector. The eigenvalues are ranked in an ascending order such that $\Theta_1$ is the smallest and $\Theta_T$ is the largest one. Equation~\eqref{eq2.1.3} facilitates removal of certain eigenvalues by setting the range of $t$ from $a$ to $b$, say,
\begin{equation}
\tilde{\Sigma}=\sum_{t=a}^{b}\Theta_tV_tV^{\dag}_t \ .
\label{eq2.1.4}
\end{equation}
In total, there are  $T=96$ eigenvalues $\Theta_t$ for $\Sigma$. Due to $K<T$, $\Sigma$ has $T-K+1=62$ zero eigenvalues and $K-1=34$ non-zero ones~\cite{Wang2022}. Using an ascending order for the eigenvalues, this means the first 62 eigenvalues are zero, i.e., $\Theta_1,\cdots,\Theta_{62}=0$, and the eigenvalues from $\Theta_{63}$ to $\Theta_{96}$ are non-zero. To remove collective effects in time, we exclude the largest eigenvalue $\Theta_{96}$. Besides, to reduce the noise, we also remove the smallest 9 non-zero eigenvalues from $\Sigma$, such that the non-zero eigenvalues remained in $\Sigma$ is from $\Theta_{72}$ to $\Theta_{95}$. Thus, we set $a=72$ and $b=95$ to obtain a reduced-rank covariance matrix $\tilde{\Sigma}$, which is well-defined as demonstrated in Refs.~\cite{Wang2020,Heckens2020}. We finally acquire a reduced-rank correlation matrix by 
\begin{equation}
\tilde{D}=\tilde{\sigma}^{-1} \tilde{\Sigma}  \tilde{\sigma}^{-1}\ ,
\label{eq2.1.5}
\end{equation}
where $\tilde{\sigma}$ is the diagonal matrix of the standard deviations
\begin{equation}
\tilde{\sigma}=\mathrm{diag} \left(\tilde{\sigma}_1,\cdots,\tilde{\sigma}_T \right)\ ,
\label{eq2.1.6}
\end{equation} 
calculated from the reduced-rank covariance matrix $\tilde{\Sigma}$. For each day, we work out a reduced-rank correlation matrix $\tilde{D}$. For the 362 days with complete data in 2015, we obtain 362 matrices $\tilde{D}$ in total.

\subsection{Quasi-stationary states }
\label{sec22}

Let $\tilde{D}(i)$ be the reduced-rank correlation matrix for day $i$ and $\tilde{D}_{t_kt_l}(i)$ be the elements of $\tilde{D}(i)$ at times $t_k$ and $t_l$. A similarity measure between the correlation structures of two days is quantified by
\begin{equation}
\eta_{ij}=\big\langle | \tilde{D}_{t_kt_l}(i)-\tilde{D}_{t_kt_l}(j) |\big\rangle_{t_kt_l} \ ,
\label{eq2.1.7}
\end{equation}
where $|\cdots|$ stands for the absolute value and $\langle \cdots \rangle_{t_kt_l}$ represents the average over all matrix elements. The numbers $\eta_{ij}$ yields a $N\times N$ similarity matrix $\eta$, where $N=362$ as 362 days are available in 2015. Furthermore, we define a distance between two observations, i.e., reduced-rank correlation matrices, as a squared Euclidean distance
\begin{equation}
E_{ij}=\sum\limits_{n=1}^{N}(\eta_{in}-\eta_{jn})^2 \ .
\label{eq3.4.2}
\end{equation}
If two reduced-rank correlation matrices $\tilde{D}(i)$ and $\tilde{D}(j)$, respectively, show similarities with other matrices, these two matrices are very likely to have similar structures, leading to a short distance~\cite{Wang2020}. Using the squared Euclidean distance matrix, we classify 362 reduced-rank correlation matrices with $k$-means clustering~\cite{Lloyd1982}. The optimal number of clusters is 6 due to the minimal standard deviation of the averages of intra-cluster distances~\cite{Pharasi2018}. The $k$-means clustering is a centroid-based clustering method that recurrently assigns each observation to the cluster of its closest centroid until the cluster to which each observation belongs does not change or iterations reach the preset maximal number, respectively. In comparison with other clustering methods, it has the advantage of low complexity and high computing efficiency. 

\begin{figure}[tbp]
\begin{center}
\includegraphics[width=1\textwidth]{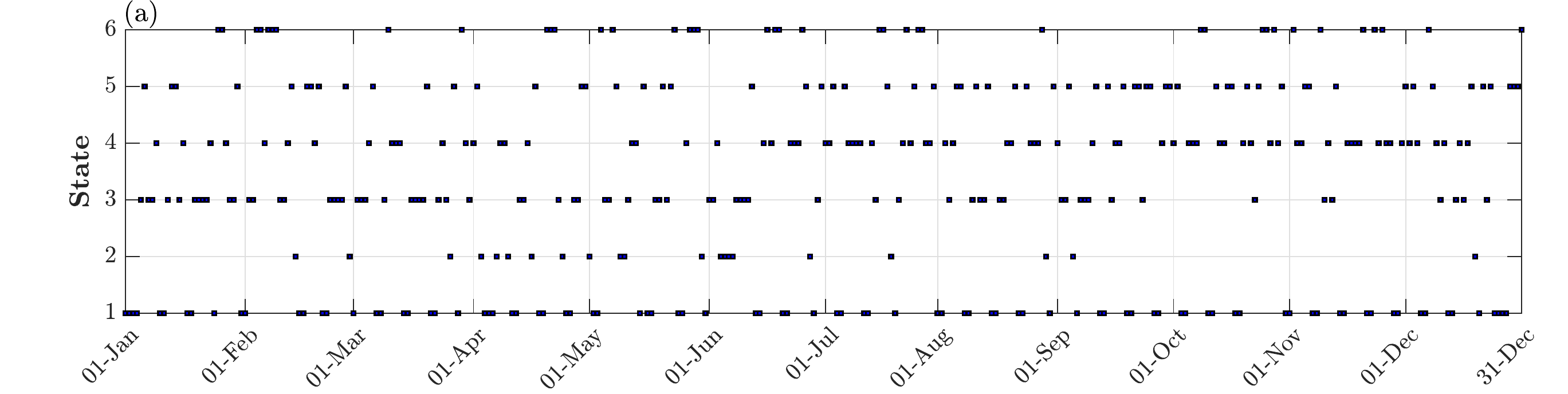}
\includegraphics[width=1\textwidth]{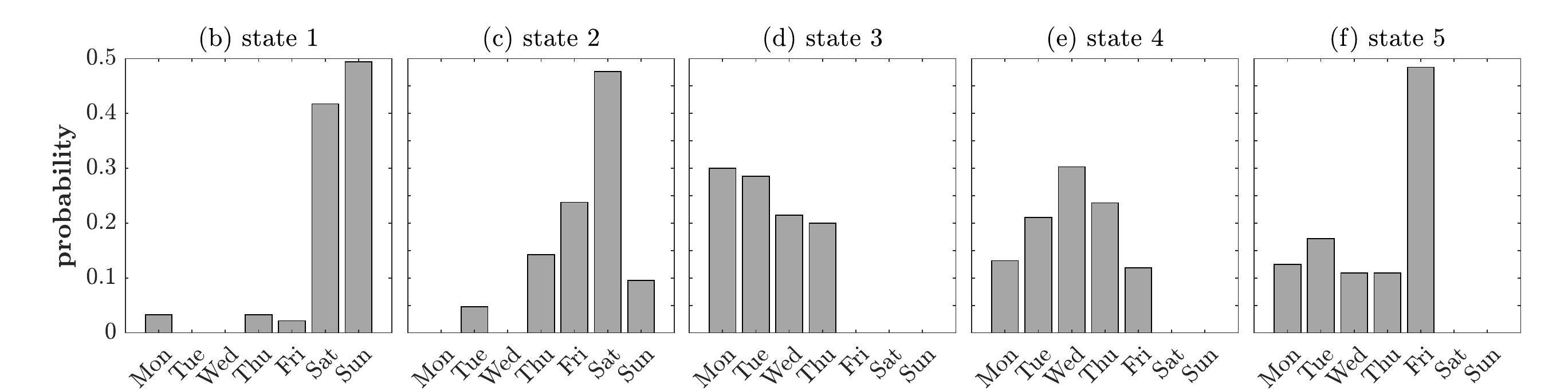}
\caption{(a) The time evolution of the traffic quasi-stationary states, where the days with the missing data are classified as state 6. (b--f) The probability distribution for every weekday on which a given state occurs.}
\label{fig1}
\end{center}
\end{figure}

We refer to a cluster in the set of reduced-rank correlation matrices as a quasi-stationary state. Therefore, the six clusters yield six quasi-stationary states in the temporal correlation structure of the traffic flow. For more details of identifying the quasi-stationary states, we refer the reader to Ref.~\cite{Wang2020}. A time evolution of each state and the probability for every weekday that a given state occurs are displayed in Fig.~\ref{fig1}. As explained in Ref.~\cite{Wang2020}, state 1 is a holiday state composed of $97\%$ weekends and public holidays, while states 3, 4 and 5 are workday states containing $100\%$ workdays. To be more specific, state 3 spans from Monday to Thursday, state 4 from Monday to Friday and state 5 is dominated by Friday data. Furthermore, state 2 is a mixing state that contains public holidays, weekends and workdays. Since the missing values in correlation matrices  obstruct the similarity measurements and thus the affected correlation matrices cannot be used in the clustering algorithm, we manually assign all correlation matrices with missing values to an additional cluster, i.e. cluster 6, which results in state 6. Within state 6, the proportion of missing values to all values in each correlation matrix jumps from $2.07\%$ to $90.87\%$.

\section{Transitions between quasi-stationary states}
\label{sec3}

We now investigate the transition between quasi-stationary states as identified in Sec.~\ref{sec2}. In Sec.~\ref{sec31}, we describe the salient features of the PageRank algorithm as well as a power method for solving Google matrices in order to obtain PageRank scores. In Sec.~\ref{sec32}, we work out the transition probabilities for quasi-stationary states, and we visualize the state transition networks for the four quarters of 2015. We then identify the dominant states for the traffic system with the PageRank algorithm in Sec.~\ref{sec33}, and explore the influence of random events on the traffic system in Sec.~\ref{sec34}. In Sec.~\ref{sec35}, we further analyze the number of transition steps which are typically taken until recurrence of a state.

\subsection{PageRank algorithm}
\label{sec31}

PageRank~\cite{Brin1998,Page1999,Langville2011} is a well known algorithm for measuring the importance of website pages via their link structure. Recently, it has been used to evaluate the importance of nodes from directed graphs, to recommend important nodes, to make link prediction between two nodes and so on~\cite{Gleich2015}. Here, we use it to measure the importance of quasi-stationary states based on their transition probability matrices. 

The PageRank algorithm~\cite{Brin1998,Page1999,Langville2011,Wills2006} contains three important matrices: a hyperlink matrix $H$, a modified hyperlink matrix $S$ and a Google matrix $G$. If there are $n$ webpages, considering each webpage as a node, the connections among these nodes can form a directed web graph, formulated by a $n\times n$ hyperlink matrix $H$. If webpage $i$ has $l_i$ ($l_i\geq 1$)  links to other webpages, the probability of each link that a random surfer selects to reach webpage $j$ is $H_{ij}=1/l_i$, where $H_{ij}$ is an element of $H$. When the random surfer arrives at a dangling webpage without any link, the probability that one reaches other webpages by selecting links is zero, i.e. $H_{ij}=0$. In this case, one can enter an address on a web browser to visit other webpages. The probability to choose any of the $n$ webpages is $1/n$ rather than zero. Therefore, the hyperlink matrix $H$ is modified to a new matrix $S$ by 
\begin{equation}
S=H+\delta w \ , 
\label{eq3.1.1}
\end{equation}
where $\delta$ is a column vector with the elements $\delta_i=1$ for $l_i=0$ and $\delta_i=0$ otherwise, and $w$ is a uniform row vector $w=[1/n~1/n~\cdots~1/n]$. Let a scalar $\alpha$ ($0\leq \alpha<1$) as a damping factor, indicating the probability that the random surfer is guided by the matrix $S$ to move. Besides, there is also a probability of $1-\alpha$ for the surfer to randomly move to next webpage by entering an address on a web browser, regardless of the links in the current webpage. Therefore, the matrix $S$ is modified to a Google matrix $G$,
\begin{equation}
G=\alpha S + (1-\alpha)\mathbb{1}\zeta \ .
\label{eq3.1.2}
\end{equation}   
Here $\mathbb{1}$ is a column vector of ones and $\zeta$ is a row probability distribution vector, named personalization vector. The hyperlink structure of the web is weighted heavily when $\alpha$ is close to one, but is lost when $\alpha=0$. Brin and Page chose $\alpha=0.85$ with regard to the weight of the hyperlink structure and the rate of convergence of the power method~\cite{Brin1998,Page1999}. They also assigned a uniform vector $\zeta=[1/n~1/n~\cdots~1/n]$ to account for the probability that the random surfer views any of the $n$ webpages without selecting the links. 

The Google matrix $G$ is an irreducible square matrix as the random surfer can proceed to any webpage from any other webpage either by selecting links or by entering an address on a web browser. Each element of $G$ indicates the probability from one webpage to another webpage and is between 0 and 1. The sum of the elements in each row is equal to one. Let $\pi=[\pi_1\cdots \pi_n]$ be a row probability distribution vector, where $\pi_n$ is the $n$-th component of this vector. The row probability distribution vector $\pi$ is a stationary distribution when it fulfills $\pi G=\pi$~\cite{Cover1999}, leading to a unique solution for this equation. One way to estimate the stationary distribution $\pi$ is the power method~\cite{Bronson2008,Langville2011,Golub2013}. A eigenvalue with the largest magnitude is referred to as the dominant eigenvalue of a matrix. The eigenvector corresponding to the dominant eigenvalue is named dominant eigenvector. The power method is a technique for approximating to the dominant eigenvector of a square matrix. Given an initial vector $\pi^{(0)}$, which can be assigned to a uniform row vector $[1/n 1/n \cdots 1/n]$ with $n$ elements, the power method performs successive iterative calculations by
\begin{equation}
\pi^{(m)}=\pi^{(m-1)} G=\pi^{(0)}  G^m \ .
\label{eq3.1.3}
\end{equation} 
When $m$ tends to be very large such that $\pi^{(m)}=\pi^{(m-1)}$, the above equation results in a stationary distribution $\pi=\pi^{(m)}$. As mentioned, the elements in $G$ are between 0 and 1 and the sum of the elements in each row of $G$ is equal to 1. Due to these characteristics, the Google matrix $G$ is a row stochastic matrix having the fact that the dominant eigenvalue, i.e., the largest magnitude among all eigenvalues of $G$, is equal to one~\cite{Haveliwala2003}. Thus, the vector $\pi$ can be viewed as the dominant eigenvector of $G$. Each element of $\pi$ as an eigenvector component is termed a PageRank score. The PageRank score of a webpage stands for the probability that a random surfer is more likely to visit this webpage based on the link structure of the web. Hence this score estimates the importance of a webpage in a PageRank sense. The most important webpage has the largest probabilities and also the highest PageRank score for the random surfer to visit. 

\subsection{Transition probabilities of quasi-stationary states}
\label{sec32}

To estimate the likelihood of a state transition, we define a transition probability $p_{ij}^{(\tau)}$ from state $i$ to state $j$ with $\tau$ steps as the number $n_{ij}^{(\tau)}$ of state $i$ changing over to state $j$ normalized by the total number of state $i$ changing over to all states during a given time period $\mathcal{T}$ through $\tau$ steps,
\begin{equation}
p_{ij}^{(\tau)}=\mathrm{Pr}\Big(X(t+\tau)= j | X(t)= i\Big)=\frac{n_{ij}^{(\tau)}}{\sum\limits_{q=1}^{\mathcal{K}}n_{iq}^{(\tau)}} \ ,
\label{eq3.2.1}
\end{equation}
where $X(t)$ represents a state at time $t$ and $\mathcal{K}$ indicates the total number of quasi-stationary states in the traffic system. Here $\mathcal{K}=6$. For a given time period and fixed $\tau$ steps of transition, $\sum_{j=1}^{\mathcal{K}}p_{ij}^{(\tau)}=1$. Since the quasi-stationary state identified on a day-to-day time scale, $\tau$ steps of state transition reflect that a state changes over to another state after $\tau$ days.

\begin{figure}[tb]
\begin{center}
\includegraphics[width=1\textwidth]{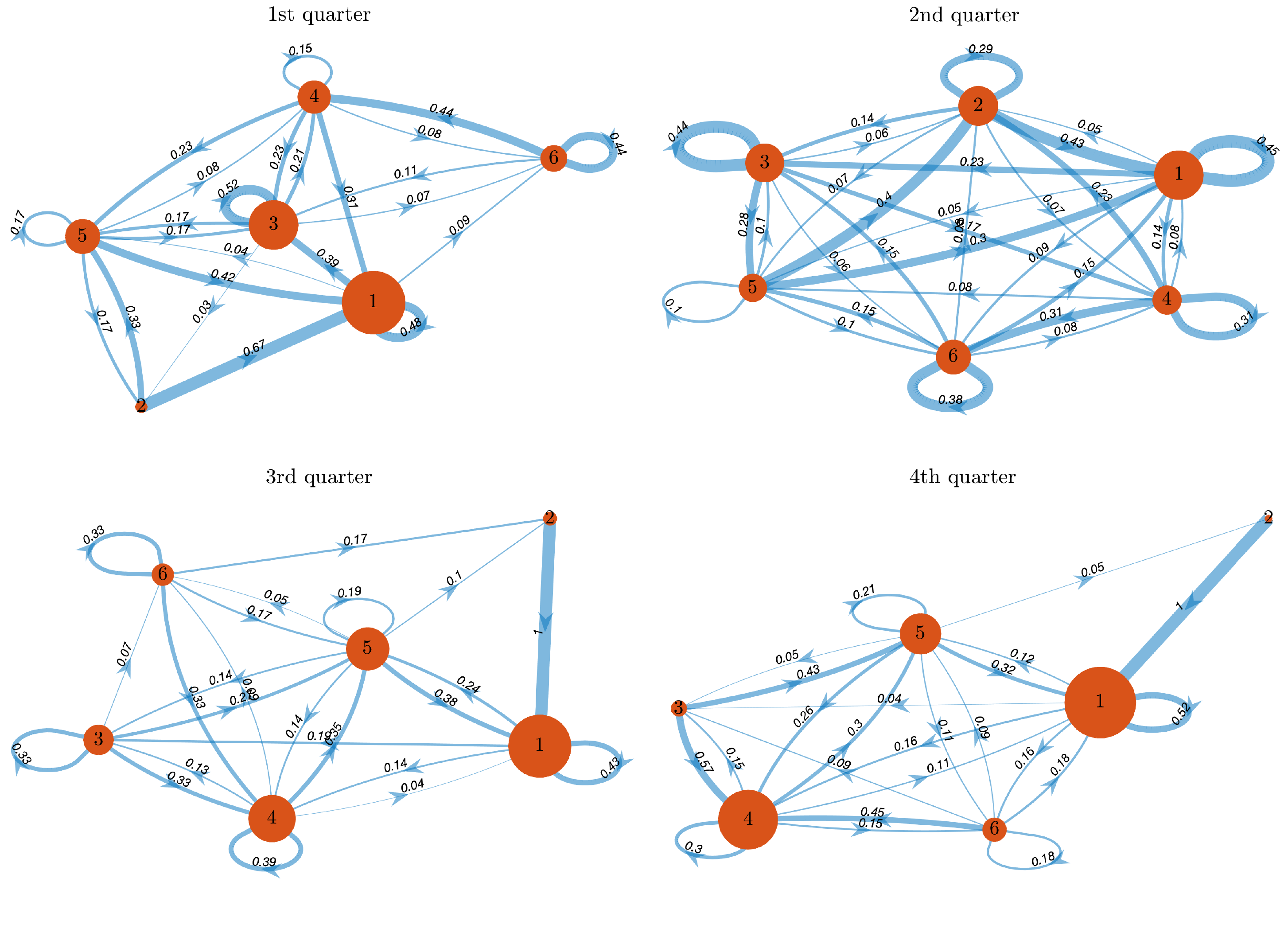}
\vspace*{-0.5cm}
\caption{The transition networks of six quasi-stationary states during each quarter of 2015. The red nodes labeled by numbers represent the states, the size of the nodes indicates the sum of probabilities that other states change over to this state, the width of a blue line between two nodes visualizes the size of a transition probability between two states, and the arrow on each blue line indicates the direction of transition from one state to another state.}
\label{fig2}
\end{center}
\end{figure}

For each quarter of 2015, we work out the transition probabilities $p_{ij}^{(1)}$  through one-step transition, i.e. $\tau=1$ day.  With the entries $p_{ij}^{(1)}$, we obtain a $6\times 6$ transition matrix $P^{(1)}$. From the transition matrix, we construct a directional network among quasi-stationary states, i.e., a state transition network. Each node of the network stands for a quasi-stationary state. Each edge between two nodes is weighted by a transition probability between two states. If the transition probability is zero, an edge is absent between two nodes. The direction of an edge indicates the direction of the transition from a state to another state. The size of a node visualizes the sum of incoming probabilities from other states to this state, i.e., $\sum_i p_{ij}^{(1)}$. As shown in Fig.~\ref{fig2}, the differences in network structures for four quarters evidently reveal the changes of transition probabilities among states. For the first quarter of 2015, other states are more likely to move to state 1, i.e. to a holiday state, and to state 3, i.e. to a workday state. The traffic behavior during this period is easily understood, as people work mainly in usual hours from Monday to Thursday,  leave work early on Friday, and rest on weekends. For the second quarter, state 2 also becomes significant for the traffic system. The dominance of state 2, which occurs on workdays and holidays, suggests that the traffic behavior is changed by some reasons, e.g., the increase of outdoor activities due to weather condition. For the third and the fourth quarters, the transition from other states to state 1 and to states 4 and 5, i.e. to two workday states, are more frequent in the traffic system. In particular, the workday state that frequently occurs shifts from state 3 during the first two quarters to states 4 and 5 in the current two quarters. This implies a large event, e.g. a road construction, makes a difference on the behavior of commuter flow on the motorways. Moreover, state 2 only changes over to state 1 and almost disappears during these two quarters. The changed traffic behavior has an impact on the temporal correlation structures of traffic flow, leading to a season-dependent transition network among quasi-stationary states for each quarter of 2015.

\subsection{Dominant states in quasi-stationary systems}
\label{sec33}

Other states are more likely to convert to a dominant one and thus the latter has a high possibility to occur during a time period. Which states dominate in the traffic system during the year of 2015, and how random events influence dominant states are the issues worth of study. Figure~\ref{fig2} shows that for a long time period, e.g., half or one year, the transitions among quasi-stationary states in the traffic system exhibit a non-Markovian feature. Only in a relatively short time period, may the quasi-stationary states behave in a Markov manner. We set this short time window as $\mathcal{T}=60$ days and always move this time window forward by one day. In this way, we work out a series of transition matrices $P^{(\tau)}(d)$ with the moving time windows, where the last day of each time window is labeled as the time point $d$.   

To determine the dominant state during a time period, we apply the PageRank algorithm to our transition matrices $P^{(\tau)}(d)$. To be more specific, the quasi-stationary states acquire the role of the webpages and the state transition matrix $P^{(\tau)}(d)$ the role of the hyperlink matrix $H$. Random events in a regular traffic system, e.g., traffic accidents, road construction, large events in a city or bad weather condition, are similar to the random behavior that a random surfer visits a webpage by inputting an address on a web browser instead of selecting any link on a previous webpage. For each time window $d$, the $6\times 6$ Google matrix in our case reads,
\begin{equation}
G(d)=\alpha \Big(P^{(\tau)}(d)+\delta w\Big) + (1-\alpha)\mathbb{1}\zeta \ .
\label{eq3.3.1}
\end{equation} 
Here, the column vector $\delta=(\delta_1, \cdots, \delta_6)$ with $\delta_k=1$ if the number of links to other states is zero, i.e. $l_k=0$, and $\delta_k=0$ otherwise. The uniform row probability vectors $w$ and $\zeta$ are defined as $w=[1/6~\cdots~1/6]$ and $\zeta=[1/6~ \cdots~1/6]$, respectively. A PageRank score in our case quantifies the importance of a quasi-stationary state. In contrast to others, the dominant state is the most important state with the highest PageRank score. The solution of $\pi(d) G(d)=\pi(d)$ of Eq.~\eqref{eq3.1.3} results in $\pi(d)$ including the PageRank scores of six states during the time window $d$. 

\begin{figure}[tb]
\begin{center}
\includegraphics[width=1\textwidth]{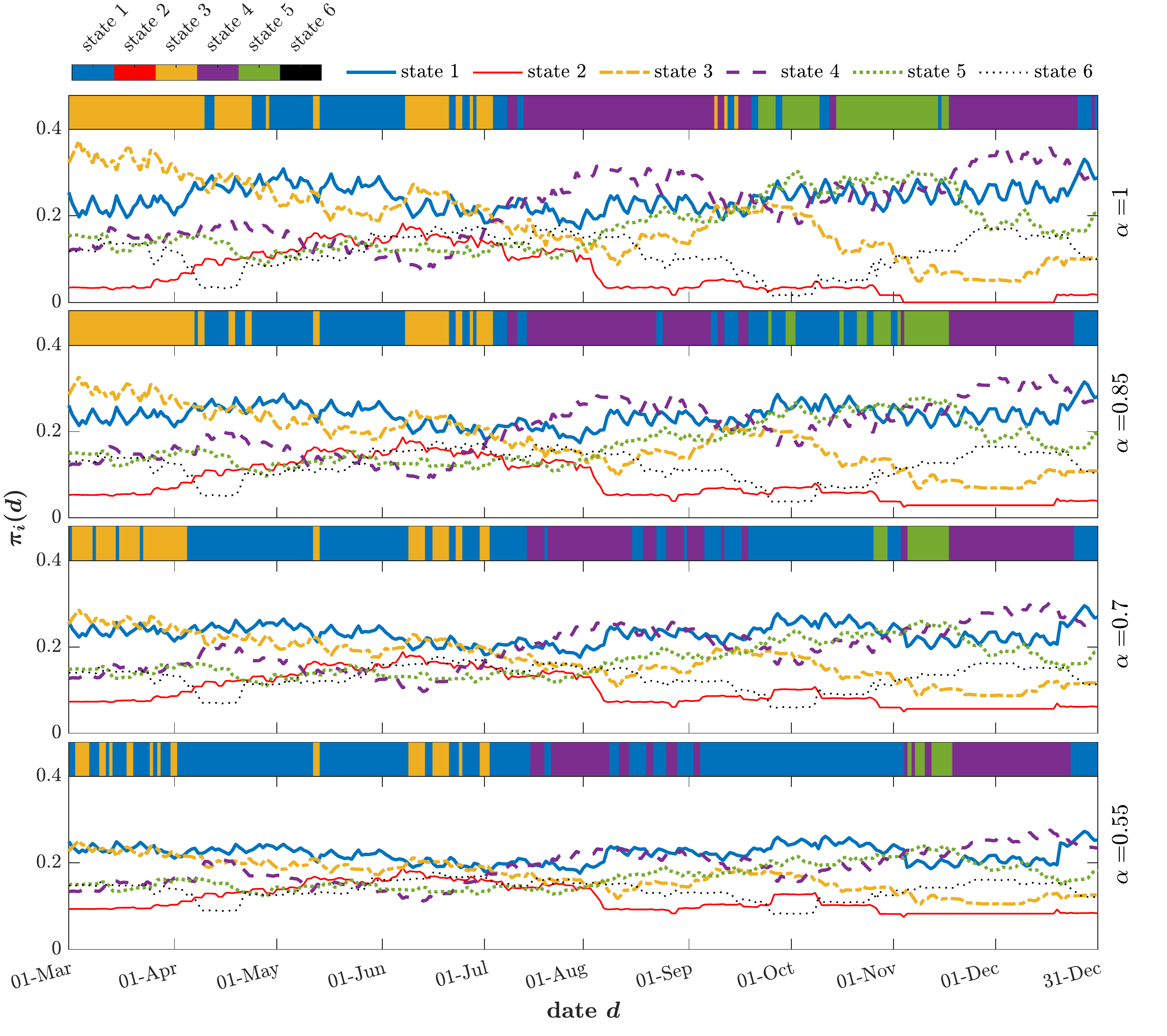}
\vspace*{-0.5cm}
\caption{The time evolution of dominant states (shown by the long color bars) and of PageRank scores $\pi_i(d)$ (shown by lines) for six quasi-stationary states with damping factors $\alpha=1,~0.85,~0.7,~0.55$ respectively. As the length of a time window of 60 days, the results of the first two months in 2015 do not show up.}
\label{fig3}
\end{center}
\end{figure}

\begin{figure}[tb]
\begin{center}
\includegraphics[width=1\textwidth]{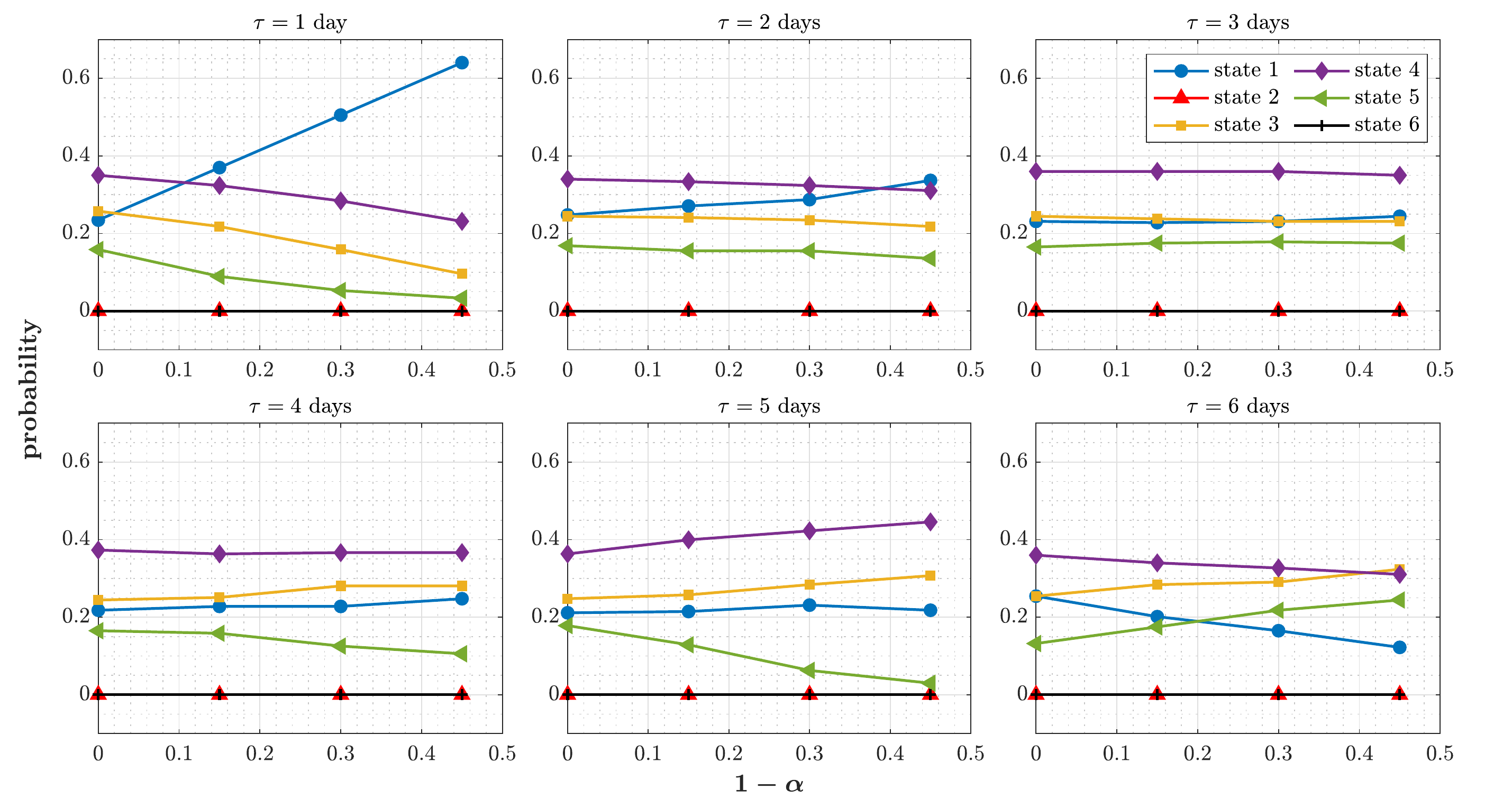}
\vspace*{-0.5cm}
\caption{Occurring probability of dominant states versus the occurring probability of random events $1-\alpha$ for the time windows of 60 days in 2015}
\label{fig4}
\end{center}
\end{figure}

\begin{figure}[tb]
\begin{center}
\includegraphics[width=1\textwidth]{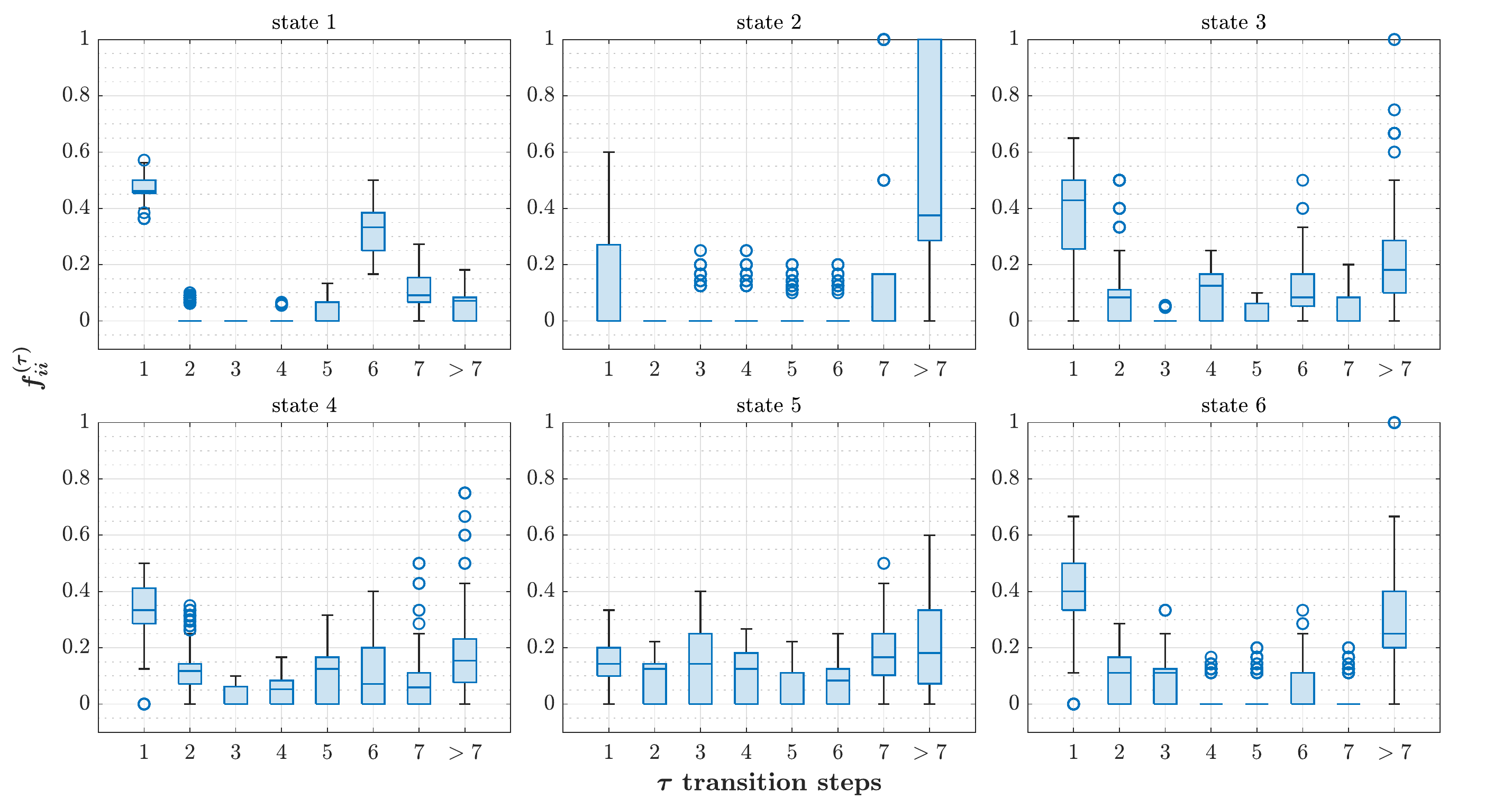}
\vspace*{-0.5cm}
\caption{Probabilities $f_{ii}^{(\tau)}$ of state $i$ for different numbers $\tau$ of transition steps with the time windows of 60 days in 2015}
\label{fig5}
\end{center}
\end{figure}

For a one-step transition, i.e. $\tau=1$ day, when the damping factor $\alpha=1$ and a column vector $\delta=(0,\cdots,0)$, indicating that any state is possible to turn to any other state, Eq.~\eqref{eq3.3.1} is solved by
\begin{equation}
\pi^{(m)}=\pi^{(0)}  \Big(P^{(1)}(d)\Big)^m \ ,
\label{eq3.3.2}
\end{equation}
with a very large $m$. Here we set $m=1000$. By this setting, we obtain purely empirical results of PageRank scores that are existent and unique for six states. For each time window $d$, the state with the largest PageRank score is the dominant state, which mainly depends on the seasons. Specifically, from January to March, state 3, i.e. a workday state which shows up from Monday to Thursday, dominates the traffic system. From April to June, state 1 prevails over other states, revealing that more people are on their vacation during this period in contrast to other time periods. From July to September, state 4 is the most important state, implying more people come back from their vacation and form regular commutes from Monday to Friday. From October to December, states 5 and 4, i.e., two workday states, predominate over the other states. It is reasonable to see the dominant holiday state, i.e., state 1, shows up at the end of December due to Christmas Day. Furthermore, the importances of six states get close to each other during June and July. During this period, the dominant states alter frequently and the PageRank score of state 2 approaches a high value as compared to the low values during other periods. This drastic change implies a big influence of events on the regular traffic system, leading to an adaptation in correlations of traffic flow. Tracing back to 2015, we find a road expansion of the motorway A3 between Cologne-M\"ulheim and Leverkusen began from June, 2015~\cite{news}. This event to some extent decreased the traffic capacity on the corresponding motorway sections and changed the distribution of traffic flow in the Cologne orbital motorways.

\subsection{The influence of random events on traffic systems}
\label{sec34}

To explore the influence of random events, such as the above-mentioned road construction on  motorway A3 in 2015, on the general traffic behavior, we reduce the damping factor $\alpha$ from 1 to 0.85, 0.7 and 0.55, respectively. The reduction of the damping factor corresponds to an increase of the occurring probability $1-\alpha$ of random events for the traffic system. As seen in Fig.~\ref{fig3}, the increase of random events prolongs the time period dominated by the holiday state while shrinking the time period dominated by the workday states. Recall the coronavirus breakout in 2020 in Germany. During the year of 2020, most people were encouraged to work at home or advised to be quarantined at home~\cite{Corona2020}. A similar example is related to the period with extreme weather condition~\cite{Weather2021}, e.g., heavy rain, tornadoes and flooding. Such extreme weather condition restrict most people to indoor activities. Accordingly, the heavy traffic burdens on workdays are relieved by these random events and the traffic system behaves as the one on holidays. 

The last conclusion applies to the state transition with one step. What if the transition between states takes two or more steps? To study this, we work out the probability of dominant states versus the probability $1-\alpha$ of random events under different steps of transition, i.e. $\tau=1,2,\cdots,6$ days. The occurring probability of a dominant state is the occurring number of a state dominating the traffic system during a whole year over the total occurring number of all dominant states in the traffic system during a whole year. For one-step transition, the trends of occurring probabilities of dominant states in Fig.~\ref{fig4} quantitively corroborate the qualitatively assessment of the time span that a dominant state occupies in Fig.~\ref{fig3}. For two- to four-step transitions, the trends of occurring probabilities of dominant states keep constant in general, regardless of the increasing probabilities of random events. However, after five- to six-step transitions, a decrease and an increase are separately evident for state 5. It seems that the increase of random events at the initial time makes it less possible to find the state 5 as a dominant state after five days, but more possible after six days. Such contradictory behavior also can be seen for state 1. As mentioned, it becomes more likely for state 1 to dominate the traffic system after one-step transition when increasing the random events. However, state 1 as a dominant state becomes unlikely after six-step transition. We can image that for a random event, i.e. a winter storm, occurring at one day, the load of motorways lessens on the next day since many people prefer to work at home or ask for a leave due to this event. Such situation will change on the day after next day or next several days, but is less possible to persist as long as six days, because of measures taken by the administration. 

\subsection{The recurrence of quasi-stationary states}
\label{sec35}

The above study focuses on the transition from one state to the same or a different state after $\tau$ steps. In the following, we will restrict the transition between the same states and explore the likelihood that after a certain number of steps a state first recurs. 

We define the probability that state $i$ first returns to itself in a $\tau$-step transition
\begin{equation} 
f_{ii}^{(\tau)}=\mathrm{Pr}\Big(X(t+\tau)= i, X(t+\tau')\neq i \mathrm{~for~} \tau'=1,\cdots,\tau-1 | X(t)= i\Big)=\frac{n_{ii}^{(\tau)}}{\sum\limits_{\tau=1}^{\infty}n_{ii}^{(\tau)}} \ ,
\end{equation}
such that $\sum_{\tau=1}^{\infty} f_{ii}^{(\tau)}=1$. Here, $n_{ii}^{(\tau)}$ is the number of the first recurrence of state $i$ by a $\tau$-step transition. Figure~\ref{fig5} shows the probability $f_{ii}^{(\tau)}$ for each state with each step $\tau$. State 1 is a holiday state that presents mostly on Saturdays and Sundays. Hence, the first recurrence of state 1 is more often after one- or six-step transition. State 2 is a state comprised of workdays and holidays and only becomes slightly important during a special period around June in Fig.~\ref{fig3}. Therefore, state 2 usually first returns to itself after more than one week in the sense of one year. States 3 and 4 are only for workdays such that there is a high likelihood for them to first recur after one-step transition. Despite of the fact that state 5 contains only the workdays, it presents on Fridays more than on other workdays. As a result, it reappears either after one week or more than one week. We cluster all correlation matrices with missing values into state 6. These missing values are attributed to the failure of inductive loop detectors or some events, such as road construction, which prevent the detection of traffic data. State 6 first recurs mostly after one step and often after more than seven steps of transition. This reveals that if we find the detection failure of traffic data, we are more likely to find this failure again after one day. We may also find this failure after more than one week, but the frequency is not as high as we find it after one day. Comparing with Fig.~\ref{fig1} to infer the number of steps that a state recurs for the first time, Fig.~\ref{fig5} quantifies the probability of each number of steps for the first recurrence of each state, in particular of states 2 and 6.

\section{Conclusions}
\label{sec4}

Using Cologne orbital motorways as an example, we studied the transition between quasi-stationary states in the temporal correlation structure of traffic systems. We found five quasi-stationary states representing five different structures in the reduced-rank correlation matrices of traffic flow. In the reduced-rank correlation matrices, the collective behavior and noise information in traffic are removed. We also considered one additional quasi-stationary state that included the correlation matrices with missing values. The total six quasi-stationary states form a time series of states in 2015. For this series, we therefore worked out the  transition probabilities between states. 

The constructed state transition networks derived from transition probability matrices reveal a seasonal dependence of states. Specifically, the traffic system is more likely to get into state 1, i.e. a holiday state, and state 3, i.e. a workday state, during the first two quarters, but into state 1 and state 4, i.e. a workday state, during the last two quarters in 2015. Moreover, state 2, i.e., a state containing both workdays and holidays, occur more frequently during the second quarter of 2015.  

The most important state is the state that dominates at a time period with a high possibility to occur in a traffic system. Utilizing the PageRank algorithm, we identified the importances of six quasi-stationary states with a time window of 60 days, which shifted forward by one day.  The dominant states once more reveal a seasonal dependence. For one-step transitions, the increase of occurring probability of random events simulated by the PageRank algorithm leads the traffic system to behave as the one on holidays more than on workdays. On the contrary, for the transition with six steps, the holiday state dominates the traffic system less and less with the increase of random events.  

The number of transition steps for the first recurrence of a quasi-stationary state is closely related to the characteristics of this state. The holiday state, i.e., state 1 that appears mostly on Saturdays and Sundays, highly probably first returns to itself by one step or six steps of transition, while the two workday states, i.e., states 3 and 4 that often occur from Mondays to Fridays, first recur frequently after one day. Another workday state, i.e., state 5 that emerges frequently on Fridays, typically recurs after a longer period of at least a week. The first recurrence of state 6 gives an estimate for the occurrence of failure in the inductive loop detectors. It seems that there is typically a malfunctioning on two different time scales, either after one day or after a period of more than a week.

The empirical study of transitions of quasi-stationary states helps to disclose hidden information in a traffic system that traditional methods do not reveal. Such information, e.g., the emergence or recurrence of a quasi-stationary state, might in turn clarify traffic behavior and operational modes as well as the spatial or temporal development of correlation patterns.

\section*{Acknowledgements}
We gratefully acknowledge funding via the grant ``Korrelationen und deren Dynamik in Autobahnnetzen'', Deutsche Forschungsgemeinschaft (DFG, 418382724). We thank Strassen.NRW for providing the empirical traffic data. We also thank Sebastian Gartzke for fruitful discussions.

\section*{Author contributions}
T.G. and M.S. proposed the research. S.W. and T.G. developed the methods of analysis. S.W. performed all the calculations. All authors contributed equally to analyzing the results, writing and reviewing the paper.

\end{document}